\documentclass[reprint, aps, prl, showpacs, showkeys]{revtex4-1}

\usepackage{graphicx}
\usepackage{dcolumn}
\usepackage{bm}

\begin{document} 

\title{Femtosecond response time measurements of a $Cs_{2}Te$ photocathode}

\author{A. Aryshev}
\email[Corresponding author: ]{alar@post.kek.jp}
\affiliation{KEK: High Energy Accelerator Research Organization,1-1 Oho, Tsukuba, Ibaraki 305-0801, Japan}
\author{M. Shevelev}
\affiliation{KEK: High Energy Accelerator Research Organization,1-1 Oho, Tsukuba, Ibaraki 305-0801, Japan}
\author{Y. Honda}
\affiliation{KEK: High Energy Accelerator Research Organization,1-1 Oho, Tsukuba, Ibaraki 305-0801, Japan}
\author{N. Terunuma}
\affiliation{KEK: High Energy Accelerator Research Organization,1-1 Oho, Tsukuba, Ibaraki 305-0801, Japan}
\author{J. Urakawa}
\affiliation{KEK: High Energy Accelerator Research Organization,1-1 Oho, Tsukuba, Ibaraki 305-0801, Japan}

\begin{abstract}

Success in design and construction of a compact high-brightness accelerator system is strongly related to the production of ultra-short electron beams. Recently the approach to generate short electron bunches or pre-bunched beams in RF guns directly illuminating a high quantum efficiency semiconductor photocathode with femtosecond laser pulses becomes attractive. The measurements of the photocathode response time in this case are essential. With a novel approach of the interferometer-type pulse splitter deep integration into commercial Ti:Sa laser system used for RF gun it has become possible to generate pre-bunched electron beam and obtain continuously variable electron bunch separation. In combination with a well-known zero-phasing technique it allows to directly estimate the response time of the most commonly used $Cs_{2}Te$ photocathode. It was demonstrated that peak-to-peak rms time response of $Cs_{2}Te$ is of the order of $370$ fs and thereby it is possible to generate and control a THz sequence of relativistic electron bunches by a conventional S-band RF gun. This result can also be applied for investigation of another cathodes material, electron beam temporal shaping and further opens a possibility to construct wide--range tunable, table--top THz FEL.

\end{abstract}


\maketitle

The importance of further development of the new high-brightness compact relativistic electron sources is rapidly growing. The main optimization factors along with charge and stability are operational lifetime and short temporal response of the photocathodes as well as low intrinsic emittance of the generated electron beam \cite{bazarov, dowel, mari}. The improvements of the photocathode RF gun technology have made a significant impact for X-ray FELs \cite{xfel, emma}, electron diffraction facilities \cite{ed} novel light sources and future colliders design \cite{piney, tenen}. Also it has become crucial technology for table--top accelerator--based X-ray and coherent radiation sources for medical \cite{medic} and biological \cite{biol} use. The generation of a high-brightness THz-frequency coherent radiation pulses (when the radiation intensity is proportional to the number of particles per bunch squared \cite{coher}) is strongly connected to generation of short electron bunches. 

Usually to obtain such conditions different bunch compression methods are used \cite{chiad, ding}. However this requires significant beamline space allocation and notably increases overall facility cost. That is why consideration of a simpler approach to generate short electron bunches or even pre-bunched  electron beams directly illuminating a photocathode with femtosecond duration laser pulses become attractive. The application of such a beam is very wide since it can be used to generate an intense spectrum-tunable THz pulses, give enhancement to radiation emission \cite{arbel, li} and also can be used for resonant excitation of the wakefields in plasma and dielectric wakefield accelerators \cite{kallos, andon}. The important properties of this pre--bunched beam could be continuously variable time separation between micro bunches, square envelope and amplitude modulation possibility within the micro-pulse sequence. However implementation of these adjustabilities sets significant limits on the RF gun drive laser pulse stacking method. Previously reported schemas were based on usage of a compact birefringent crystal arrays \cite{crys1, crys2} or interferometer--type stacking. In reality, the crystal--based approach when the input pulse is decomposed in two orthogonally polarized pulses with a time separation proportional to the crystal length does not make continuous time separation control possible due to a limited number of the crystal's length. While interferometer--based schemes are usually more mechanically advanced and requires tighter alignment, they can provide necessary pulse control \cite{2nfold, lyu}.  

Photocathode properties are usually dictating RF gun laser system output pulse energy, harmonic, polarization and pulse duration. Over the last decade many photocathode materials were tested. Nowadays the main choice of material revolves around a few options: metals ($Cu$, $Mg$), positive electron affinity (PEA) semiconductors, like $Cs_{2}Te$, $Cs_{3}Sb$, $CsK_{2}Sb$ and negative electron affinity (NEA) semiconductors, like $GaAs$. Typically metals have long life-time, short time-response, high damage threshold, but low quantum efficiency (Q.E.). NEA photocathodes are operated in visible spectrum, have high Q.E., but sensitive for oxidation hence require ultra-high vacuum operation and have a long response time. PEA photocathodes, especially $Cs_{2}Te$ also have a good resistance to laser damage and as it is usually reported \cite{boss, kong} they have a comparatively shorter response time of the order of picoseconds whereas there is no principle limitations for it not to be rather fast \cite{clenden}. Previously reported monte-carlo simulation indicated that the expected $Cs_{2}Te$ photocathodes response time should be around $400$~fs~\cite{ferrini} which was indirectly confirmed by the production of electron beam in the blow-out regime with the same type photocathode in a RF gun~\cite{piot}.  

Space--charge dominated electron beam properties are expected throughout initial acceleration in the RF gun \cite{seraf} hence the minimum bunch duration and intra--bunch time separation will be determined by the photocathode response time, accelerating gradient, laser spot size on the photocathode and laser pulse duration. 

Here we demonstrate for the first time that, the peak-to-peak rms time response of the $Cs_{2}Te$ photocathode is of the order of $370$ fs, and thereby it is possible to generate and control a THz sequence of a few hundred femtoseconds relativistic electron bunches by a conventional S-band RF gun. The result further widens the potential of designing a truly table-top THz FEL based on super-radiant undulator \cite{arbel} or Smith-Purcell radiation \cite{spr}.  

\section{Results and Discussion}
\begin{figure}[h!]
\includegraphics{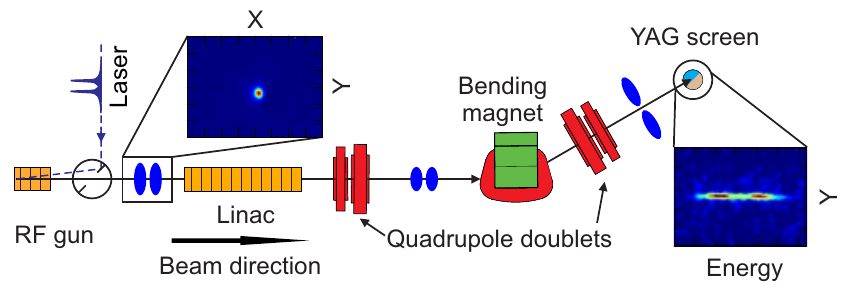}
\caption{LUCX beamline schematic.}
\label{fig3}
\end{figure}

The measurements presented in this paper were done at the Laser Undulator Compact X-ray facility (LUCX) at High Energy Accelerator Research Organization (KEK). LUCX schematic diagram is shown in Fig. \ref{fig3}. It consists of the $3.6$-cell S-band RF gun \cite{abhay} and the normal conductivity compact linear accelerator \cite{fukuda}. Two klystrons are used for the RF gun and accelerating structure. The Titanium-Sapphire laser system was employed for femtosecond LUCX operation mode. The electrons are generated by a photocathode illuminated by a train of two laser pulses (generated with a pulse divider, see the Methods section) with a variable delay in a range from a few picoseconds to a total overlap, resulting in generation of two separate electron bunches. 

The two micro--bunches are accelerated up to 7 MeV in the RF gun and energy chirped by linear accelerator. The phase of the gun is set at $25$ deg. with accelerating gradient $80$ MV/m. As the linac generates a time--energy correlation in the electron beam, the two bunches have different energies at the end of the accelerator and when dispersed by a bending magnet the different time slices of the electron bunches are projected to different transverse position on the $300\;\mu$m--thick $0.5\%$ $Ce$-doped yttrium aluminium garnet ($YAG$) luminescent screen located in the beamline section with $250$ mm horizontal dispersion. This is usually referred to as zero--phasing technique \cite{wang1} and commonly used for rms electron bunch length measurements. However it requires a careful calibration and knowledge about initial beam parameters \cite{seeman}. Also, it is important to notice that the single electron bunch length estimation from the given measurements rather reflect variation of the longitudinal phase-space ellipse caused by RF gun beam dynamics than just represent photocathode time response. In this respect the measurements based on usage of a two micro--bunches with rms length and separation range much smaller than the RF wavelength $\lambda_{RF}$ can give more accurate insight on photocathode emission temporal properties as long as bunches phase-space ellipses are not overlapped.

\begin{figure}[t!]
\includegraphics{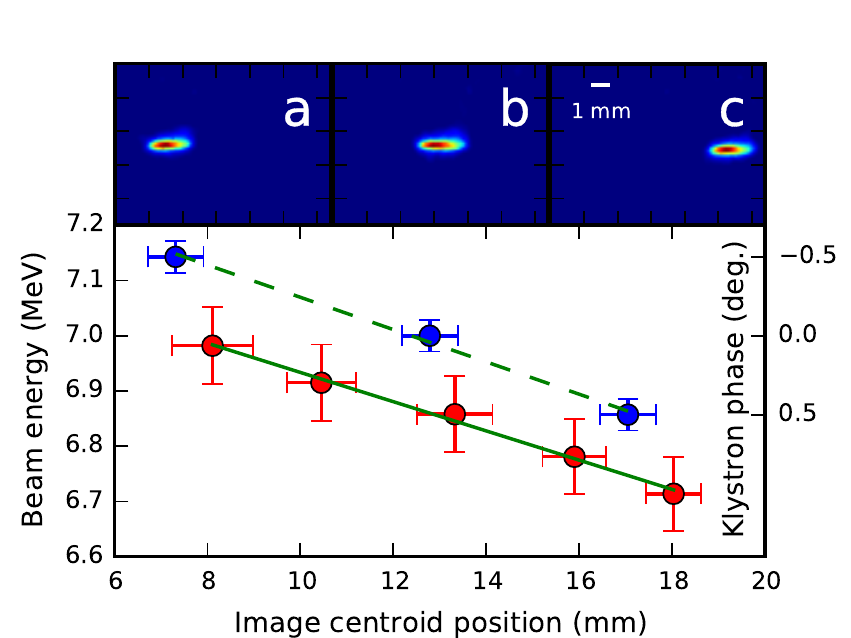}
\caption{Top row: typical electron density distributions for different bending magnet current. Bottom: beam image centroid position vs electron beam energy (red dots) with linear fit: amplitude $7.2\pm0.02$MeV and slope $2.65\cdot10^{-2}\pm1.16\cdot10^{-3}$ MeV/mm; Beam image centroid position vs RF phase (blue dots) with linear fit: slope $0.102\pm7.3\cdot10^{-3}$ deg./mm}
\label{fig4}
\end{figure}

\begin{figure}[]
\includegraphics{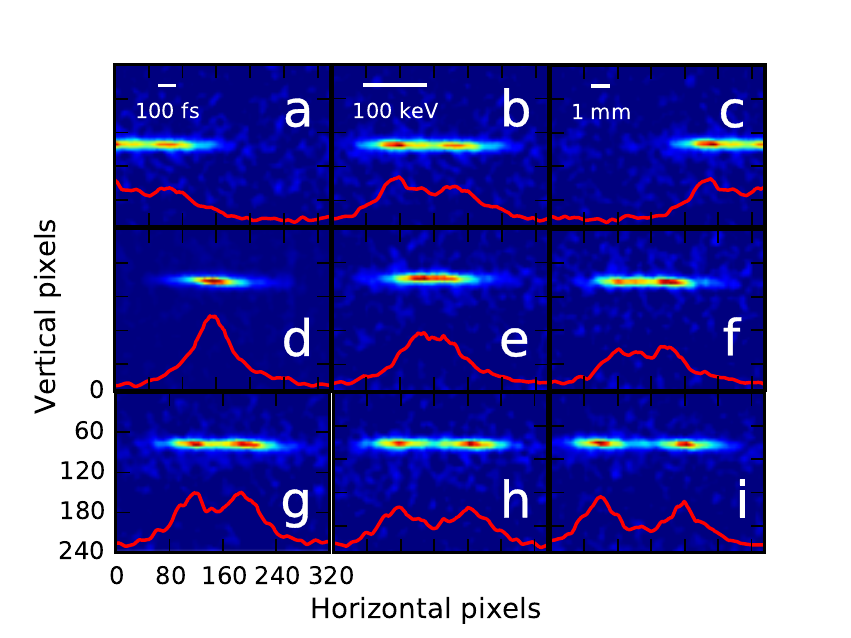}
\caption{Typical electron density distribution measured for (a): $-0.25$ deg., (b): $0$ deg., (c): $0.25$ deg. accelerating phase; 
(d - i): Typical two micro--bunch electron density distribution measured for $0$ deg. accelerating phase and (d): $4.8$ mm, (e): $4.815$ mm, (f): $4.825$ mm, (g): $4.850$ mm, 
(h): $4.865$ mm, (i): $4.875$ mm relative $M2$ mirror positions.}
\label{fig5}
\end{figure}

Initial energy calibration performed without energy chirp for each micro--bunch charge $1$pC shows that the averaged energy of the two micro--bunches is equal to $E = 7.17\pm0.002$ MeV, energy spread is $dE = 24.88\pm1.0$ keV, and hence $dE/E \sim0.33\%$, Fig.\ref{fig4}. The RF zero--crossing point is determined by finding the phase that results in no transverse image centroid movement at the spectrometer screen when the cavity is turned on and off with the same bending magnet setting, Fig.\ref{fig4} (b) and Fig.\ref{fig5} (b).   

To obtain the time calibration the correlation of the RF phase with image centroid shift on the screen was measured, Fig.\ref{fig4} blue dots and Fig.\ref{fig5} (a), (b), (c). The linear approximation of this correlation gives the scale of the horizontal image size in RF degrees, which is recalculated to the time scale as follows. The linear slope of the calibration is $0.102\pm7.3\cdot10^{-3}$ deg./mm, which is $99.20\pm7.09$ fs/mm assuming $1$ deg. RF is $972.6$ fs. The experimental parameters in this measurement are $\lambda_{RF}$ of $10.5$ cm (S--band RF, $2856$ MHz) and linac RF amplitude $V_{RF}$ of about $44$ MV. Other estimated beam parameters are rms normalized transverse emittance $\epsilon_{x}$ of $0.3$ $\pi$ mm mrad, rms longitudinal emittance of $0.15$ $\pi$ keV mm, and rms transverse beam size of $0.1 \times 0.1$ mm at the zero--phasing cavity, respectively. Since higher--order horizontal focusing terms caused by energy deviations as assumed from beamline optics are four orders of magnitude smaller than the dispersion term, the maximum slope of the longitudinal phase-space ellipse $dE/dz$ is $4$ times smaller than the RF slope $2\pi q V_{RF} /\lambda_{RF}$ of $2.6$ GeV/m. The screen monitor is placed in a location where the dispersive beam size contribution to the measured profile $\eta_{x}dE/E = 750$ $\mu$m is larger than the betatron size contribution $[\beta_{x}\epsilon_{x}/\gamma]^{1/2}= 200$ $\mu$m, where $\gamma$, $\beta_{x}$, $dE/E$ and $\eta_{x}$ are the Lorentz factor, horizontal beta function, energy spread and horizontal dispersion respectively. The $YAG$ screen image broadening effect was neglected since the electron beam charge was low \cite{lump_inj}. The rms resolution of approximately $30$ fs for this experimental arrangement is limited by the transverse beam size. It should be stressed that the measured electron beam distribution and length obtained by this measurement are valid not for the transverse profile measuring location but for the zero--phasing cavity entrance. Moreover it linearly depends on initial longitudinal electron distribution near the photocathode. Hence two micro--bunch minimal energy (and hence time) separation can be considered as the cathode response time estimate as it only depends on the laser pulse separation and is limited by the cathode response time, space-charge effect in the RF gun (which affects both the transverse and longitudinal beam profiles) and zero--phasing energy spectrometer resolution.
\begin{figure}[t!]
\includegraphics{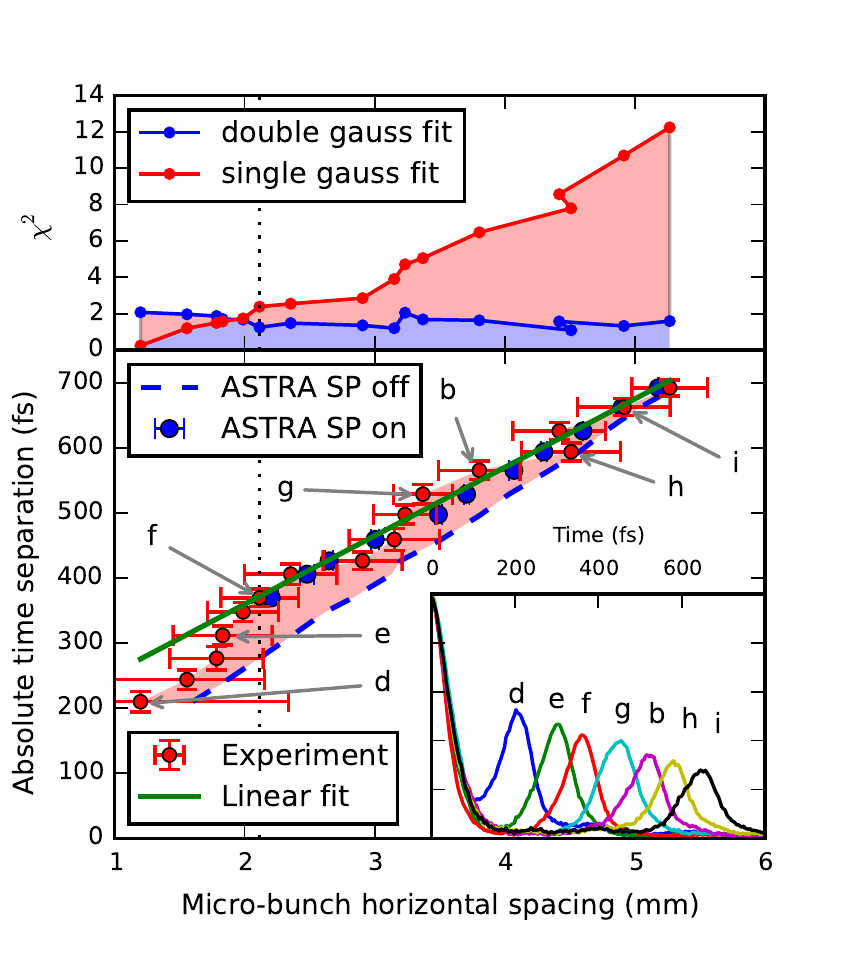}
\caption{Top: Comparison of reduced $\chi^{2}$ for single- and double- gaussian fitting of the experimental data. Bottom: Dependence of micro-bunch horizontal spacing and micro-bunch time separation with linear fit (green line): slope $107.95\pm9.4$ fs/mm. Dashed blue line and scatter blue dots are ASTRA simulation performed for space--charge off (SP off) and space--charge on (SP on) respectively. Inset: laser micro-pulse cross-correlations. Markers corresponds to Fig.\ref{fig5}}
\label{fig6}
\end{figure}
To prove this statement the additional mechanical calibration method was developed. The method is based on one micro-bunch arrival time change and recording the bunch--to--bunch distance change on $YAG$ screen while the linac RF phase is set to zero--crossing phase. The sequence of screen images shows gradual increase of the energy difference between the two bunches, Fig.\ref{fig5} (d - i). Simultaneously the fundamental harmonic laser pulses cross-correlations were acquired to confirm laser micro--pulse spacings, Fig.\ref{fig6} inset. Both the screen image vertical projection and laser cross--correlations were fitted by double Gaussian function and the correlation of peak separations were plotted changing $M2$ mirror position (see the Methods section). A $100$ consecutive measurements were acquired for each data point. As can be seen, this technique gives similar scale factor of $107.95\pm9.4$ fs/mm for $YAG$ screen horizontal image size, that is consistent with phase calibration within standard deviations. In addition the simulation of experimental condition by ASTRA tracking package \cite{astra} was done, Fig.\ref{fig6}. It shows a noticeable space--charge effect contribution (dashed blue line and blue scatters) and well--matched with experimental data. Top part of the Fig.\ref{fig6} represents the comparison of the mean square weighted deviation (i.e. reduced chi-squared $\chi^{2}$) for single- and double- Gaussian fitting of the experimental data.

Peak--to--peak rms separation as small as $90.5$ keV or $369.48\pm27$ fs have been obtained in the experiment which is supported by simulation and different fitting functions comparison. Lower data points show gradual peaks overlap and better fitted by the single Gaussian function. Although the energy resolution in this experiment is limited by the  horizontal dispersion at the $YAG$ screen, it was estimated that the maximum achievable separation is limited to a few percent by chromatic effects in the electron beam transport system. 

Presented two micro--bunch method demonstrates a highly efficient way to directly estimate $Cs_{2}Te$ photocathode response time which can also be applied for investigation of another cathodes material used to generate electron beam in RF guns, electron beam temporal shaping, resonant excitation of the wakefields in plasma and dielectric wakefield accelerators. It is also important to emphasis that measurements were done for acceleration field gradient of $80$ MV/m and a thin $Cs_{2}Te$ layer of $10$ nm what in fact is the most typical usage condition of this type photocathodes in RF gun. The dependence of the response time on Q.E. of this type photocathode remains an open question for further work. To the best of our knowledge, presented result is an absolute record among all photocathode materials ever used in RF guns reported up to the date. In addition, for the first time the deep integration of the laser pulse-divider into commercial Titan-Sapphire laser system has been shown. This result further opens a possibility to construct wide–range tunable compact THz FEL. Taking into account the absolute performance record among all known photocathode materials, we are confident that our contribution will have a high impact in the field of radiation physics, compact accelerator based high brightness THz source development and photo-injectors design. 

\section{Methods}
\subsection{Response time}
By the cathode temporal response we understand the emission time shaped by the stochastic smearing due to scattering processes of the emitted electrons. This effect lengthens the extracted pulses and produces long tails and hence the single electron bunch minimal duration or minimal temporal separation of two and more electron bunches will be limited by the cathode response time.

\subsection{Photocathode preparation}
The cathode preparation system is well described in \cite{teru}. The system is similar to many RF gun photocathode preparation systems that are commonly used in many institutes worldwide, CERN, LASA, DESY, Fermilab, LBNL, etc.  The photocathode was formed on the fine diamond powder polished molybdenum substrate \cite{cheval}. The evaporation process includes a $\sim10$ nm  $Te$ film deposition on the substrate surface with subsequent activation by $Cs$ until the Q.E. reaches a few percent level. However due to $Cs$ quick diffusion and oxidation the resulting Q.E. was stabilized at a $0.5\%$ level. This result was confirmed by both xenon lamp and measurements with an electron beam. In the first case the $266$ nm photon beam from xenon lamp with diffractometer hits the cathode. By measuring light intensity and photo-current it is possible to obtain the Q.E. value with high precision. The second case is similar except that the photocathode is installed into RF gun and illuminated by the Ti:Sa laser third harmonic ($266$ nm) photons and number of emitted electrons is measured by an Inductive Current Transformer located downstream of the gun. 

\subsection{Laser splitting techniques}
\begin{figure}[h!]
\includegraphics{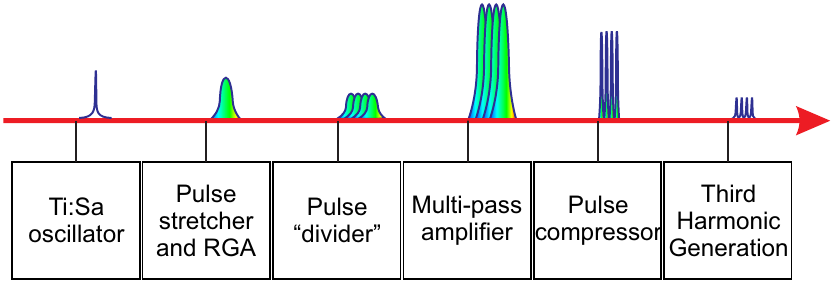}
\caption{Chirped pulse amplification with micro-pulse extension. RGA - regenerative amplifier.}
\label{fig1}
\end{figure}

\begin{figure}
\includegraphics{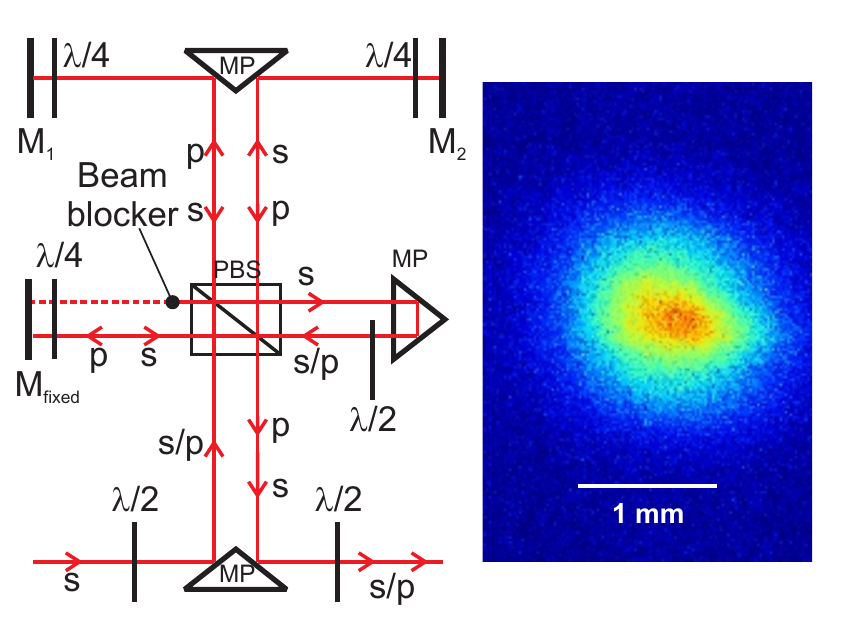}
\caption{Pulse divider general scheme (left) and typical $500\times500\;\mu$m rms UV profile at virtual cathode (right). $PBS$ - polarized beam splitter, $MP$ - prism mirror, $M1$ and $M2$ motorized mirrors, $M_{fixed}$ fixed mirror, $\lambda/2$ and $\lambda/4$ half- and quarter- wave retardation plates respectively.}
\label{fig2}
\end{figure}

To generate micro-bunch sequence the Titanium-Sapphire laser system was chosen, Fig.\ref{fig1}. Short pulses of a few tens of femtoseconds from the oscillator are temporarily stretched to a picosecond level before they are amplified in a regenerative amplifier (RGA) up to a few $\mu$J. The ``pulse divider'' (PD) was introduced right after the RGA so the pulses are split and recombined with controllable delay by the double--pass Michelson--like interferometer, Fig.\ref{fig2} (left), where linear polarizations are denoted as $s$ and $p$, and $s/p$ represents a mixed polarization states. This minor modification allows the generation of a spectrally chirped picosecond pulses sequence with variable time separation. 
After that they are amplified up to $2$ mJ per micro--pulse by the multi--pass Ti:Sa amplifier and re-compressed back to a few tens of femtoseconds, typically $45$ fs . This is possible due to the micro--pulses having the same polarization after the multi--pass amplifier. As a result $1.3$ mJ of each micro--pulse energy at the Ti:Sa fundamental harmonic ($800$ nm) was available at the laser system output which is sufficient for third harmonic ($266$ nm) generation that is required for the $Cs_{2}Te$ cathode as well as providing a large energy margin to increase number of micro--bunches. The laser system was also extended to allow direct measurement of the generated micro--pulse durations and its time separations by the method based on the registration of the second harmonic ($400$ nm) energy cross distribution produced in the nonlinear crystal \cite{kolm1}. Resulting cross--correlations give calibrated \cite{salin} absolute temporal measurements which is then compared with electron beam measurements. In case of micro--pulsed input the cross--correlation dependence has additional peaks symmetric around the main correlation. The time separation between main correlation and the satellite peak is equal to the real time separation between micro--pulses. The third harmonic diagnostics includes laser spot imaging (Fig.\ref{fig2}, right) at the distance equivalent to the cathode (so--called ``virtual cathode'') and pulse energy measurements. In the current configuration, when first input to the fixed mirror is blocked, PD can produce two micro pulses with time separation proportional to the pass difference of the interferometer arms. The minimum UV pulse duration at the photocathode after transport through a $11$~m long optical line and passage through UV telescope and a vacuum window is estimated as $130$ fs due to group delay dispersion \cite{gdd}.
The only one movable mirror $M2$ determines time separation between pulses and the relative time offset is proportional to two times its displacement. The $25$ nm spectral bandwidth of each picosecond laser pulse when separated by more than $70$ fs ensures no longitudinal interference between micro-pulses \cite{int} that in turn gives uniform pulse intensities within the micro--train unlike the results presented in \cite{wang2, fiorito}.

\begin{acknowledgments}
The authors would like to thank S.Araki, M.Fukuda, P.Karataev, L. Sukhikh and other colleagues from AGTaX collaboration for their valuable help, useful discussions and support. This work was supported by Photon and Quantum Basic Research Coordinated Development Program from the Ministry of Education, Culture, Sports, Science and Technology, Japan, JSPS KAKENHI grant numbers $23226020$ and $24654076$ and Leverhulme Trust Network ``Advanced Research on Generation of THz and X-ray Radiation'' (IN – 2015 – 012).
\end{acknowledgments}

\section{Contributions}
A.A. and M.S. commissioned Ti:Sa laser system, developed and integrated pulse divider, developed electron beam optics and run experiments; M.S. performed the ASTRA simulations; Y.H. provided insights to optimize zero-phasing technique. A.A. analyzed data and wrote the manuscript. N.T. and J.U. provided management and oversight to the project. All authors commented on the manuscript and agreed on the contents.

\section{Competing interests}
The authors declare no competing financial interests.

\end{document}